\documentclass[aps,prl,superscriptaddress,showpacs,floatfix,twocolumn]{revtex4}
\usepackage{graphicx}	% Include figure files

\begin{document}


\title{Enhancement of the dielectron continuum 
in Au+Au collisions at $\sqrt{s_{NN}}$=200 GeV}

\newcommand{\abilene}{Abilene Christian University, Abilene, TX 79699, U.S.}
\newcommand{\banaras}{Department of Physics, Banaras Hindu University, Varanasi 221005, India}
\newcommand{\bnl}{Brookhaven National Laboratory, Upton, NY 11973-5000, U.S.}
\newcommand{\caucr}{University of California - Riverside, Riverside, CA 92521, U.S.}
\newcommand{\cns}{Center for Nuclear Study, Graduate School of Science, University of Tokyo, 7-3-1 Hongo, Bunkyo, Tokyo 113-0033, Japan}
\newcommand{\colorado}{University of Colorado, Boulder, CO 80309, U.S.}
\newcommand{\columbia}{Columbia University, New York, NY 10027 and Nevis Laboratories, Irvington, NY 10533, U.S.}
\newcommand{\dapnia}{Dapnia, CEA Saclay, F-91191, Gif-sur-Yvette, France}
\newcommand{\debrecen}{Debrecen University, H-4010 Debrecen, Egyetem t{\'e}r 1, Hungary}
\newcommand{\elte}{ELTE, E{\"o}tv{\"o}s Lor{\'a}nd University, H - 1117 Budapest, P{\'a}zm{\'a}ny P. s. 1/A, Hungary}
\newcommand{\fsu}{Florida State University, Tallahassee, FL 32306, U.S.}
\newcommand{\gsu}{Georgia State University, Atlanta, GA 30303, U.S.}
\newcommand{\hiroshima}{Hiroshima University, Kagamiyama, Higashi-Hiroshima 739-8526, Japan}
\newcommand{\ihepprot}{IHEP Protvino, State Research Center of Russian Federation, Institute for High Energy Physics, Protvino, 142281, Russia}
\newcommand{\illuiuc}{University of Illinois at Urbana-Champaign, Urbana, IL 61801, U.S.}
\newcommand{\isu}{Iowa State University, Ames, IA 50011, U.S.}
\newcommand{\jinrdubna}{Joint Institute for Nuclear Research, 141980 Dubna, Moscow Region, Russia}
\newcommand{\kaeri}{KAERI, Cyclotron Application Laboratory, Seoul, South Korea}
\newcommand{\kek}{KEK, High Energy Accelerator Research Organization, Tsukuba, Ibaraki 305-0801, Japan}
\newcommand{\kfki}{KFKI Research Institute for Particle and Nuclear Physics of the Hungarian Academy of Sciences (MTA KFKI RMKI), H-1525 Budapest 114, POBox 49, Budapest, Hungary}
\newcommand{\korea}{Korea University, Seoul, 136-701, Korea}
\newcommand{\kurchatov}{Russian Research Center ``Kurchatov Institute", Moscow, Russia}
\newcommand{\kyoto}{Kyoto University, Kyoto 606-8502, Japan}
\newcommand{\labllr}{Laboratoire Leprince-Ringuet, Ecole Polytechnique, CNRS-IN2P3, Route de Saclay, F-91128, Palaiseau, France}
\newcommand{\lawllnl}{Lawrence Livermore National Laboratory, Livermore, CA 94550, U.S.}
\newcommand{\losalamos}{Los Alamos National Laboratory, Los Alamos, NM 87545, U.S.}
\newcommand{\lpc}{LPC, Universit{\'e} Blaise Pascal, CNRS-IN2P3, Clermont-Fd, 63177 Aubiere Cedex, France}
\newcommand{\lund}{Department of Physics, Lund University, Box 118, SE-221 00 Lund, Sweden}
\newcommand{\muenster}{Institut f\"ur Kernphysik, University of Muenster, D-48149 Muenster, Germany}
\newcommand{\myongji}{Myongji University, Yongin, Kyonggido 449-728, Korea}
\newcommand{\nagasaki}{Nagasaki Institute of Applied Science, Nagasaki-shi, Nagasaki 851-0193, Japan}
\newcommand{\newmex}{University of New Mexico, Albuquerque, NM 87131, U.S. }
\newcommand{\nmsu}{New Mexico State University, Las Cruces, NM 88003, U.S.}
\newcommand{\ornl}{Oak Ridge National Laboratory, Oak Ridge, TN 37831, U.S.}
\newcommand{\orsay}{IPN-Orsay, Universite Paris Sud, CNRS-IN2P3, BP1, F-91406, Orsay, France}
\newcommand{\pnpi}{PNPI, Petersburg Nuclear Physics Institute, Gatchina, Leningrad region, 188300, Russia}
\newcommand{\riken}{RIKEN, The Institute of Physical and Chemical Research, Wako, Saitama 351-0198, Japan}
\newcommand{\rikjrbrc}{RIKEN BNL Research Center, Brookhaven National Laboratory, Upton, NY 11973-5000, U.S.}
\newcommand{\rikkyo}{Physics Department, Rikkyo University, 3-34-1 Nishi-Ikebukuro, Toshima, Tokyo 171-8501, Japan}
\newcommand{\saispbstu}{Saint Petersburg State Polytechnic University, St. Petersburg, Russia}
\newcommand{\saopaulo}{Universidade de S{\~a}o Paulo, Instituto de F\'{\i}sica, Caixa Postal 66318, S{\~a}o Paulo CEP05315-970, Brazil}
\newcommand{\seoulnat}{System Electronics Laboratory, Seoul National University, Seoul, South Korea}
\newcommand{\stonybrkc}{Chemistry Department, Stony Brook University, Stony Brook, SUNY, NY 11794-3400, U.S.}
\newcommand{\stonycrkp}{Department of Physics and Astronomy, Stony Brook University, SUNY, Stony Brook, NY 11794, U.S.}
\newcommand{\subatech}{SUBATECH (Ecole des Mines de Nantes, CNRS-IN2P3, Universit{\'e} de Nantes) BP 20722 - 44307, Nantes, France}
\newcommand{\tenn}{University of Tennessee, Knoxville, TN 37996, U.S.}
\newcommand{\titech}{Department of Physics, Tokyo Institute of Technology, Oh-okayama, Meguro, Tokyo 152-8551, Japan}
\newcommand{\tsukuba}{Institute of Physics, University of Tsukuba, Tsukuba, Ibaraki 305, Japan}
\newcommand{\vandy}{Vanderbilt University, Nashville, TN 37235, U.S.}
\newcommand{\waseda}{Waseda University, Advanced Research Institute for Science and Engineering, 17 Kikui-cho, Shinjuku-ku, Tokyo 162-0044, Japan}
\newcommand{\weizmann}{Weizmann Institute, Rehovot 76100, Israel}
\newcommand{\yonsei}{Yonsei University, IPAP, Seoul 120-749, Korea}
\affiliation{\abilene}
\affiliation{\banaras}
\affiliation{\bnl}
\affiliation{\caucr}
\affiliation{\cns}
\affiliation{\colorado}
\affiliation{\columbia}
\affiliation{\dapnia}
\affiliation{\debrecen}
\affiliation{\elte}
\affiliation{\fsu}
\affiliation{\gsu}
\affiliation{\hiroshima}
\affiliation{\ihepprot}
\affiliation{\illuiuc}
\affiliation{\isu}
\affiliation{\jinrdubna}
\affiliation{\kaeri}
\affiliation{\kek}
\affiliation{\kfki}
\affiliation{\korea}
\affiliation{\kurchatov}
\affiliation{\kyoto}
\affiliation{\labllr}
\affiliation{\lawllnl}
\affiliation{\losalamos}
\affiliation{\lpc}
\affiliation{\lund}
\affiliation{\muenster}
\affiliation{\myongji}
\affiliation{\nagasaki}
\affiliation{\newmex}
\affiliation{\nmsu}
\affiliation{\ornl}
\affiliation{\orsay}
\affiliation{\pnpi}
\affiliation{\riken}
\affiliation{\rikjrbrc}
\affiliation{\rikkyo}
\affiliation{\saispbstu}
\affiliation{\saopaulo}
\affiliation{\seoulnat}
\affiliation{\stonybrkc}
\affiliation{\stonycrkp}
\affiliation{\subatech}
\affiliation{\tenn}
\affiliation{\titech}
\affiliation{\tsukuba}
\affiliation{\vandy}
\affiliation{\waseda}
\affiliation{\weizmann}
\affiliation{\yonsei}
\author{S.~Afanasiev}	\affiliation{\jinrdubna}
\author{C.~Aidala}	\affiliation{\columbia}
\author{N.N.~Ajitanand}	\affiliation{\stonybrkc}
\author{Y.~Akiba}	\affiliation{\riken} \affiliation{\rikjrbrc}
\author{J.~Alexander}	\affiliation{\stonybrkc}
\author{A.~Al-Jamel}	\affiliation{\nmsu}
\author{K.~Aoki}	\affiliation{\kyoto} \affiliation{\riken}
\author{L.~Aphecetche}	\affiliation{\subatech}
\author{R.~Armendariz}	\affiliation{\nmsu}
\author{S.H.~Aronson}	\affiliation{\bnl}
% ...
\collaboration{PHENIX Collaboration} \noaffiliation

\date{\today}

\begin{abstract}

The PHENIX experiment has measured the dielectron continuum in 
$\sqrt{s_{NN}}$=200 GeV Au+Au collisions. In minimum bias collisions the 
dielectron yield in the mass range between 150 and 750 MeV/c$^2$ is 
enhanced by a factor of 3.4$\pm$0.2(stat.)$\pm$1.3(syst.)$\pm$0.7(model) 
compared to the expectation from our model of hadron decays. The 
integrated yield increases faster with the centrality of the collisions 
than the number of participating nucleons, suggesting emission from 
scattering processes in the hot and dense medium. The continuum yield 
between the masses of the $\phi$ and the $J/\psi$ mesons is consistent 
with expectations from correlated $c\bar{c}$ production, though other 
mechanisms are not ruled out.

\end{abstract}

	
\maketitle

%%%%%%%%%%%%%%%%%%%%%%%%%%%%%%%%%%---intro

Electron-positron pairs, or dileptons in general, have proven to be an 
excellent tool to study collisions of heavy ions at ultra-relativistic 
energies. Because leptons do not interact strongly, emission of dileptons 
from the hot matter created at RHIC should leave an imprint on the observed 
dilepton distributions. Emission from the hot matter may include thermal 
radiation and in-medium decays of mesons with short lifetimes, like the 
$\rho$ meson, while their spectral functions may be strongly modified. 
However, below the mass of the $\phi$ meson, these sources compete with a 
large contribution of e$^+$e$^-$--pairs from Dalitz decays of pseudoscalar 
mesons ($\pi^0, \eta, \eta'$) and decays of vector mesons ($\rho, \omega, 
\phi$). Above the $\phi$ meson mass up to 4.5 GeV/c$^2$, competing sources 
are dilepton decays of charmonia ($J/\psi, \psi'$) and semileptonic decays of 
$D$ and $\bar{D}$ mesons, correlated through flavor conservation, which lead 
to a continuum of masses. In addition to thermal radiation, energy loss of 
charm quarks in the medium might modify the continuum yield in this mass 
region.

%%%%%%%%%%%%%%%%%%%%%%%%%%%%%%%%%%%---previous measurements

The discovery of a large enhancement of the dilepton yield at masses below 
the $\phi$ meson mass in ion-ion collisions at the CERN SPS \cite{CER1} has 
triggered a broad theoretical investigation of modifications of properties of 
hadrons in a dense medium and of how these modifications relate to chiral 
symmetry restoration \cite{theory1}. These theoretical studies will benefit 
from the availability of more precise data from CERN \cite{NA60_rho,CER2} and 
GSI \cite{HADES}. An enhanced yield was also observed at higher masses, above 
the $\phi$ meson mass \cite{NA50}. Recent NA60 data suggest that the 
enhancement can not be attributed to decays of D-mesons but may result from 
prompt production, as expected for thermal radiation \cite{NA60_therm}.

%%%%%%%%%%%%%%%%%%%%%%%%%%%%%%%%%%%%---the phenix experiment

The PHENIX experiment at the Relativistic Heavy Ion Collider (RHIC) extends 
these measurements in a new energy regime by exploring Au+Au collisions at a 
center of mass energy of $\sqrt{s_{NN}}$=200 GeV. In this paper we present 
results from minimum bias data taken in 2004. Collisions were triggered 
and selected by centrality using beam-beam counters (BBC) and zero degree 
calorimeters (ZDC). We analyzed a sample of 8$\times 10^8$ minimum bias 
events.

%%%%%%%%%%%%%%%%%%%%%%%%%%%%%%%%%%%%%%---tracking and momentum resolution 

Electrons and positrons are reconstructed in the two central arm 
spectrometers of PHENIX \cite{NIM} using Drift Chambers (DC), located outside 
an axial magnetic field, which measure their momenta with an accuracy of 
$\sigma_p/p=0.7\%\oplus 1\%p/(\mathrm{GeV/c})$.
%%%%%%%%%%%%%%%%%%%%%%%%%%%%%%%%%%%%%%---eID (cut out and reference)
They are identified by hits in the Ring Imaging Cherenkov detector (RICH) 
and by matching the momentum with the energy measured in an electromagnetic 
calorimeter (EMCal) \cite{ppg066}. Electrons are reconstructed with an 
efficiency of $\sim$90\%, while a hadron contamination of $\sim$20\% remains.

%%%%%%%%%%%%%%%%%%%%%%%%%%%%%%%%%%%%%%--- acceptance

Each central arm covers $\left|\Delta\eta\right|\le0.35$ in pseudorapidity 
and $\pi/2$ in azimuthal angle. Because charged particles are deflected in 
the azimuthal direction by the magnetic field, the acceptance depends on the 
momentum and the charge of the particle, and also on the radial location of 
the detector component (DC, EMCal and RICH). The acceptance for a track with 
charge $q$, transverse momentum $p_{\rm T}$ and azimuthal emission angle 
$\phi$ can be described by:
\begin{equation} \label{eq:acc}
 \phi_{\rm min} \leq \phi+ q \frac{k_{\rm DC,RICH}}{p_{\rm T}} \leq \phi_{\rm max}
\end{equation}
where $k_{\rm DC}$ and $k_{\rm RICH}$ represent the effective azimuthal bend 
to DC and RICH ($k_{DC}=0.206$ rad GeV/c and $k_{RICH}=0.309$ rad GeV/c). One 
arm has $\phi_{\rm min}=\frac{-3}{16}\pi$ and $\phi_{\rm 
max}=\frac{5}{16}\pi$, the other arm $\phi_{min}=\frac{11}{16}\pi$ and 
$\phi_{max}=\frac{19}{16}\pi$. Only electrons with $p_{\rm T}\ge$~200~MeV/c 
are used in the analysis. The photon conversion probability was minimized by 
installing a helium bag between the beam pipe and the DC, reducing the 
material to $\sim$0.4\% of a radiation length.

%%%%%%%%%%%%%%%%%%%%%%%%%%%%%%%%%%%%%%---all pairs

In an event the source of any electron or positron is unknown and therefore 
all electrons and positrons are combined to pairs, like-sign and unlike-sign.
%%%%%%%%%%%%%%%%%%%%%%%%%%%%%%%%%%%%%%---combinatorial background
This results in a large combinatorial background which must be removed. The 
background is computed with a mixed event technique, which combines tracks 
from different events that have similar topology (centrality, collision 
vertex, reaction plane).

%%%%%%%%%%%%%%%%%%%%%%%%%%%%%%%%%%%%%%%---remove background pairs: 

In order to achieve the necessary accuracy, all unphysical correlations that 
arise from overlapping tracks or hits in the detectors, mostly in the RICH, 
must be eliminated, because they can not be reproduced by mixed events. If 
hits of both tracks of a pair overlap in any detector, the event is rejected. 
About 4\% of all pairs are removed by this event rejection.
%%%%%%%%%%%%%%%%%%%%%%%%%%%%%%%%%%%%%%---shape
Comparing measured like-sign pairs with the mixed combinatorial background 
shows that the mixing technique reproduces the shape within the statistical 
accuracy of the data.

%%%%%%%%%%%%%%%%%%%%%%%%%%%%%%%%%%%%%%---2sqrt

The absolute normalization of the unlike-sign combinatorial background is 
given by the geometrical mean of the observed positive and negative like-sign 
pairs $2\sqrt{N_{--}N_{++}}$, where, in principle, $N_{--}$ and $N_{++}$ are 
the measured number of like-sign pairs.
%%%%%%%%%%%%%%%%%%%%%%%%%%%%%%%%%%%%%%---like sign signal
There is a small correlated signal also in the observed like-sign pairs, 
which can occur if there are two e$^+$e$^-$--pairs in the final state of a 
meson, e.g. double Dalitz decays, Dalitz decays followed by a conversion of 
the decay photon or two photon decays followed by conversion of both photons. 
These ``cross'' pairs have small masses, typically less than the $\eta$ mass 
(550 MeV/c$^2$).

%%%%%%%%%%%%%%%%%%%%%%%%%%%%%%%%%%%%%%--- normalization to the like sign

We therefore determine $N_{--}$ and $N_{++}$ by integrating the mixed event 
distributions after they were normalized to the $7.5 \times 10^6$ like-sign 
pairs measured above 700 MeV/c$^2$. $N_{--}$ and $N_{++}$ are determined with 
an accuracy of 0.12\%.
%%%%%%%%%%%%%%%%%%%%%%%%%%%%%%%%%%%%%%---kappa correction
The normalization is multiplied by 1.004 to account for the fact that the 
event rejection removes 10\% more like-sign than unlike-sign pairs. This 
correction factor was estimated, using mixed events, with an accuracy better 
than 50\%.
%%%%%%%%%%%%%%%%%%%%%%%%%%%%%%%%%%%%%%---sys
Adding the statistical error and the uncertainty due to the event rejection 
in quadrature gives an accuracy of 0.25\% on the normalization.

%%%%%%%%%%%%%%%%%%%%%%%%%%%%%%%%%%%%%%---photon conversion

After subtraction of the combinatorial background, physical background from 
photon conversions and cross pairs is removed. Because the tracking assumes 
that the e$^+$e$^-$--pair originates at the collision vertex, pairs from 
photons that convert in or outside of the beampipe are reconstructed with 
finite mass and opening angle, which is oriented perpendicular to the 
magnetic field. A cut on the orientation of the opening angle in the field 
removes more than 98\% of the conversion pairs.

%%%%%%%%%%%%%%%%%%%%%%%%%%%%%%%%%%%%%%---cross   

Cross pairs occur as like and unlike-sign pairs. Monte Carlo simulations show 
that the rate of unlike-sign cross pairs accepted in PHENIX is 44\% of the 
rate for like-sign cross pairs. To deterime the rate of unlike-sign cross 
pairs, we scale the simulated like-sign cross pair distribution to the 
observed like-sign signal, obtained by subtraction of the mixed event 
background normalized above 700 MeV/c$^2$. We note that the like-sign signal 
is well described by the Monte Carlo simulation. The simulated unlike-sign 
cross pair distribution is scaled by the same factor and subtracted from the 
unlike-sign signal. The uncertainty of this subtraction depends on mass, but 
is $\leq$ 9\% of the final yield.

%%%%%%%%%%%%%%%%%%%%%%%%%%%%%%%%%%%%%%---RESULTS
%%%%%%%%%%%%%%%%%%%%%%%%%%%%%%%%%%%%%%---fig 1

\begin{figure*}[t]
 \includegraphics[width=0.625\linewidth]{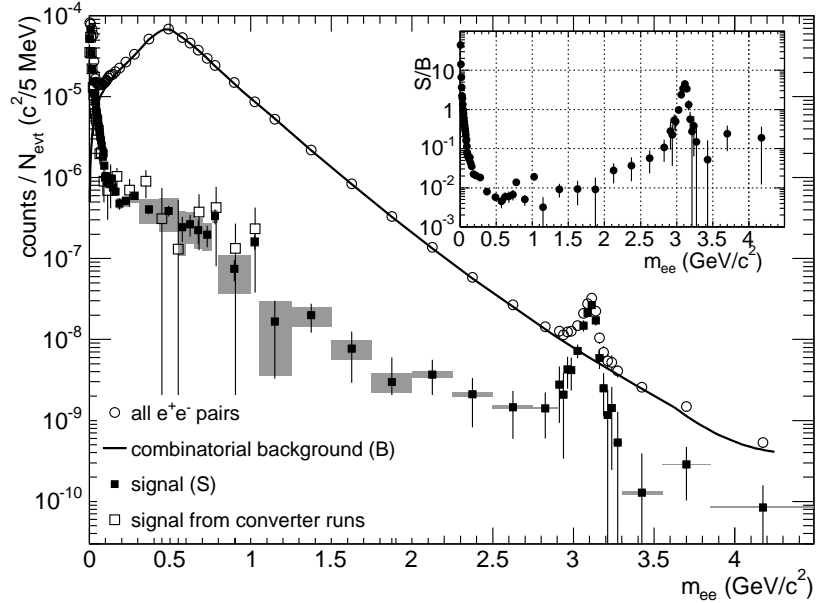}  
 \caption{\label{fig:mass}
Uncorrected mass spectra of all e$^+$e$^-$ pairs, mixed events background 
($B$) and signal ($S$) with statistical (bars) and systematic (boxes) 
uncertainties shown separately. The signal from the runs with additional 
converter is shown with statistical errors only. The insert shows the $S/B$ 
ratio. The mass range covered by each data point is given by horizontal bars.
}
\end{figure*}

Figure~\ref{fig:mass} shows the mass distribution of e$^+$e$^-$--pairs, the 
normalized mixed event background ({\it{B}}), and the signal yield ({\it{S}}) 
obtained by subtracting the mixed event background, the cross pairs and the 
conversion pairs. The insert shows the signal-to-background ratio ($S/B$). 
The systematic errors (boxes) reflect the error on the background 
subtraction, which is given by $\delta_S/S~=~0.25\%~\cdot~B/S$, added in 
quadrature to the uncertainty due to the cross pair subtraction, assumed to 
be $9\%S$ below 600 MeV/c$^2$. Despite the small $S/B$ ratio, the vector 
meson resonances $\omega$, $\phi$ and $J/\psi$ which decay directly to 
e$^+$e$^-$, and an e$^+$e$^-$--pair continuum is visible up to 4.5 GeV/c$^2$.

%%%%%%%%%%%%%%%%%%%%%%%%%%%%%%%%%%%%%%---converter

In order to check the background subtraction, a subset of data (5$\times 
10^7$ events), taken with additional material wrapped around the beam pipe to 
increase the number of photon conversions \cite{ppg066}, was analyzed. In 
this data set the combinatorial background and the cross pair contribution is 
larger by a factor of $\sim$2.5. As shown in Fig.~\ref{fig:mass}, the results 
from both data sets agree well within statistical errors, which are 30\% in 
the range from 150 to 750 MeV/c$^2$ and much less below. Considering the 
decreased $S/B$ ratio for the data with the converter we can estimate a 0.1\% 
scale uncertainty of the background normalization, well within the 0.25\% 
systematic uncertainty assigned.
\begin{figure*}[t]
 \includegraphics[width=0.625\linewidth]{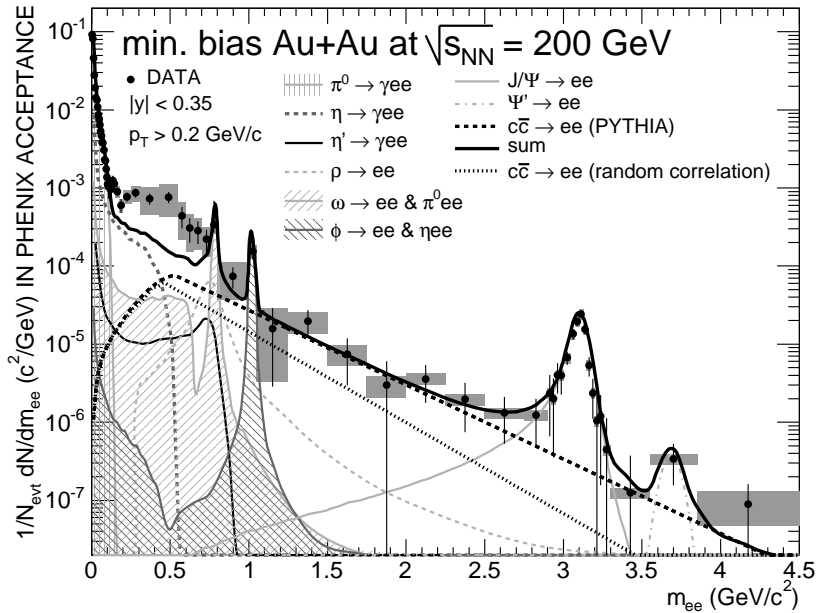}
 \caption{  \label{fig:cock}
Invariant e$^+$e$^-$--pair yield compared to the yield from the model of 
hadron decays. The charmed meson decay contribution based on PYTHIA is 
included in the sum of sources (solid black line). The charm contribution 
expected if the dynamic correlation of $c$ and $\bar{c}$ is removed is shown 
separately. Statistical (bars) and systematic (boxes) uncertainties are shown 
separately; the mass range covered by each data point is given by horizontal 
bars. The systematic uncertainty on the cocktail is not shown.}
\end{figure*}

%%%%%%%%%%%%%%%%%%%%%%%%%%%%%%%%%%%%%%---efficiency correlation

The spectra are corrected to represent the invariant yield of 
e$^+$e$^-$--pairs, with both the e$^+$ and e$^-$ in the detector acceptance 
as specified in Eq.~\ref{eq:acc}. The correction is determined using a GEANT 
simulation \cite{geant} of the PHENIX detector that includes the details of 
the detector response. Simulated e$^+$e$^-$--pairs are reconstructed with the 
same analysis chain and all cuts applied. The correction is determined double 
differentially in p$_\mathrm{T}$ and mass of the e$^+$e$^-$--pair. The 
reduction of the electron reconstruction efficiency (0.92+/-0.03) due to 
detector occupancy is corrected for.
%%%%%%%%%%%%%%%%%%%%%%%%%%%%%%%%%%%%%%---systematic error
Systematic uncertainties on the correction can be summarized as: (i) 13.4\% 
on dielectron reconstruction, which is twice the uncertainty on the electron 
reconstruction efficiency \cite{ppg066}, (ii) 6\% conversion rejection cut, 
(iii) 5\% event rejection and (iv) 3\% occupancy. These uncertainties are 
included in the final systematic error on the invariant e$^+$e$^-$--pair 
yield.

%%%%%%%%%%%%%%%%%%%%%%%%%%%%%%%%%%%%%%---fig 2
%%%%%%%%%%%%%%%%%%%%%%%%%%%%%%%%%%%%%%---cocktail

Figure~\ref{fig:cock} compares the invariant yeild to the expected yield from 
meson decays and correlated decays of charmed mesons. The cocktail of hadron 
decay contributions was estimated using PHENIX data for meson production when 
available.  As input distributions we use the measured $\pi, \eta, \phi, 
J/\psi$ yield and spectra \cite{pich, pi0, eta, ppg016, ppg068}.  For other 
mesons we use the $m_{\mathrm{T}}$ scaling procedure outlined in~\cite{ppg066}. 
The systematic uncertainties depend on mass and range from 10 to 25\%.

%%%%%%%%%%%%%%%%%%%%%%%%%%%%%%%%%%%%%%---charm

For the continuum below the $J/\psi$ the dynamic correlation of $c$ and 
$\bar{c}$ is essential, but unknown. We make two assumptions: (i) the 
correlation is unchanged by the medium and equal to what is known from p+p 
collisions. In this case we use PYTHIA \cite{pythia} scaled from the p+p 
equivalent $c\overline{c}$ cross section of 567$\pm$57$\pm$193 $\mu$barn 
\cite{ppg065} to minimum bias Au+Au collisions proportional to the number of 
binary collisions ($258 \pm 25$) \cite{pich}. We note that the p$_\mathrm{T}$ 
distribution for electrons generated by PYTHIA is softer than the spectra 
measured in p+p data but coincides with those observed in Au+Au 
\cite{ppg066}. As a second assumption (ii) there is no dynamical correlation, 
i.e. the direction of $c$ and $\bar{c}$ quarks are uncorrelated. We keep the 
overall cross section and the p$_\mathrm{T}$ distributions fixed to 
experimental data \cite{ppg066}. Other contributions from bottom and 
Drell-Yan are expected to be small in the mass region below the $J/\psi$ 
peak. Each e$^+$ and e$^-$ must fall in the PHENIX acceptance, given by 
Eq.~\ref{eq:acc}.

%%%%%%%%%%%%%%%%%%%%%%%%%%%%%%%%%%%%%%---description of fig2

The data below 150 MeV/c$^2$ are well described by the cocktail of hadronic 
sources. The vector mesons $\omega$, $\phi$ and $J/\psi$ are reproduced 
within the uncertainties. However, the yield is substantially enhanced above 
the expected yield in the continuum region from 150 to 750 MeV/c$^2$. The 
enhancement in this mass range is a factor of 3.4 $\pm$ 0.2(stat.) $\pm$ 
1.3(syst.) $\pm$ 0.7(model), where the first error is the statistical error, 
the second the systematic uncertainty of the data, and the last error is an 
estimate of the uncertainty of the expected yield. Above the $\phi$ meson 
mass the data seem to be well described by the continuum calculation based on 
PYTHIA. This is somewhat surprising, since single electron distributions from 
charm show substantial medium modifications \cite{ppg066}, and thus it is 
hard to understand how the dynamic correlation at production of the 
$c\bar{c}$ remains unaffected by the medium. A complete randomization of that 
correlation (see Fig.\ref{fig:cock}) leads to a much softer mass spectrum and 
would leave significant room for other contributions, e.g. thermal radiation.

%%%%%%%%%%%%%%%%%%%%%%%%%%%%%%%%%%%%%%---fig 3

\begin{figure}[t]
 \includegraphics[width=1.0\linewidth]{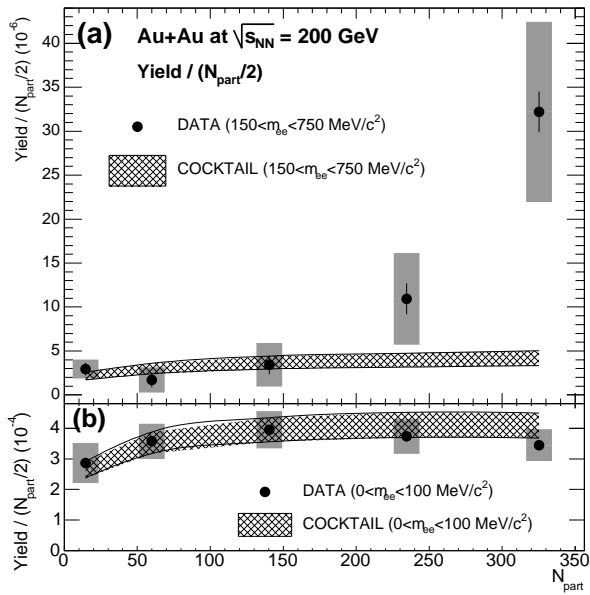}
 \caption{  \label{fig:ratio}
Dielectron yield per participating nucleon pairs ($N_{\rm part}/2$) as 
function of $N_{part}$ for two different mass ranges compared to the expected 
yield from the hadron decay model. The two lines give $\pm 1 \sigma$ 
systematic uncertainty. For the data statistical and systematic uncertainties 
are shown separately.}
\end{figure}

To shed more light on the continuum yield we have studied the centrality 
dependence of the yield in three mass windows, below 100 MeV/c$^2$, from 150 
to 750 MeV/c$^2$ and 1.2 to 2.8 GeV/c$^2$. The top panel of 
Fig.~\ref{fig:ratio} shows the centrality dependence of the yield in the mass 
region 150--750 MeV/c$^2$ divided by the number of participating nucleon 
pairs ($N_{\rm part}/2$). For comparison the yield below 100 MeV/c$^2$, which 
is dominated by low p$_\mathrm{T}$ pion decays, is shown in the lower panel. 
For both intervals the yield is compared to the same yield calculated from 
the hadron cocktail. In the lower mass range the yield agrees with the 
expectation, i.e. is proportional to the pion yield. In contrast, in the 
range from 150 to 750 MeV/c$^2$, the observed yield rises significantly 
compared to the expectation, reaching a factor of 7.7 $\pm$ 0.6(stat.) $\pm$ 
2.5(syst.) $\pm$ 1.5(model) for most central collisions. The increase is 
qualitatively consistent with the conjecture that an in-medium enhancement of 
the dielectron continuum yield arises from scattering processes like $\pi\pi$ 
or $q\bar{q}$ annihilation, which would result in a yield rising faster than 
proportional to $N_{part}$.

%%%%%%%%%%%%%%%%%%%%%%%%%%%%%%%%%%%%%%---fig 4

\begin{figure}[t]
 \includegraphics[width=1.0\linewidth]{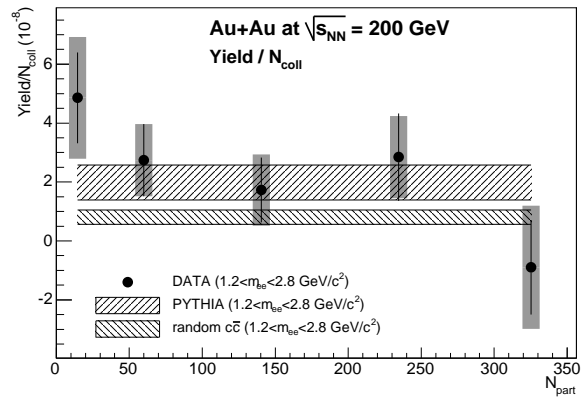}
 \caption{  \label{fig:ratio2}
Dielectron yield per number of collisions $N_{coll}$ in the mass range 1.2 to 
2.8 GeV/c$^2$ as function of $N_{\rm part}$. Statistical and systematic 
errors are shown separately. Also shown are two bands corresponding to the 
two different estimates of the contribution from charmed meson decays. The 
width of the band reflect the uncertainty of the charm cross-section only.}
\end{figure}

We normalize the yield in the mass region 1.2 to 2.8 GeV/c$^2$ to the number 
of binary collisions (Fig.~\ref{fig:ratio2}), which is the correct scaling 
for pairs from charmed meson decays \cite{ppg066}. The normalized yield shows 
no significant centrality dependence and is consistent with the expectation 
based on PYTHIA. It is also likely that a scenario where the correlation 
between the $c$ and $\bar{c}$ is randomized will require an additional 
source, e.g. a contribution from thermal radiation. This contribution could 
increase faster than linear with $N_{\rm part}$ and therefore the apparent 
scaling with $N_{coll}$ may be a mere coincidence. We note that this 
coincidence may have been observed in this mass region at the CERN SPS 
\cite{NA50}, where a major prompt component has now been suggested by NA60 
data \cite{NA60_therm}.

%%%%%%%%%%%%%%%%%%%%%%%%%%%%%%%%%%%%%%--- conclusions 

In conclusion, measurements of Au+Au collisions at $\sqrt{s_{NN}}$=200 GeV in 
the mass range 150--750 MeV/c$^2$ show a significant enhancement of the 
dielectron continuum and exhibit a clear increase with centrality of the 
collision.  The observed yield between $\phi$ and $J/\psi$ is consistent with 
the expectation from correlated $c\bar{c}$ production, but does not exclude 
other mechanisms.

% See NOTES below for information about reference citations, figures 
% and tables, using wide text for equations, landscape figures, etc.

%%%%%%%%%%%%%%%%%%%%%%%%%%%%%%%%%%%%%%%%%%%%%%%%%%%  Acknowledgements 

\begin{acknowledgments}

We thank the staff of the Collider-Accelerator and 
Physics Departments at BNL for their vital contributions.  
We acknowledge support from 
the Office of Nuclear Physics in DOE Office of Science and NSF (U.S.A.), 
MEXT and JSPS (Japan), 
CNPq and FAPESP (Brazil), 
NSFC (China), 
IN2P3/CNRS, and CEA (France), 
BMBF, DAAD, and AvH (Germany), 
OTKA (Hungary), 
DAE (India), 
ISF (Israel), 
KRF and KOSEF (Korea), 
MES, RAS, and FAAE (Russia),
VR and KAW (Sweden), 
U.S. CRDF for the FSU, 
US-Hungarian NSF-OTKA-MTA, 
and US-Israel BSF.

\end{acknowledgments}

%%%%%%%%%%%%%%%%%%%%%%%%%%%%%%%%%%%%%%%%%%%%%%%%%%%%%%%%%%%  References 

\end{document}